\documentclass[12pt]{article}
\usepackage{jheppub}

\usepackage[english]{babel}
\usepackage{amsmath,amssymb}
\numberwithin{equation}{section}

\def\r{\rho}
\def\pa{\partial}
\def\dalpha{\dot{\alpha}}
\def\dbeta{\dot{\beta}}
\def\dgamma{\dot{\gamma}}
\def\dkappa{\dot{\kappa}}
\def\dtau{\dot{\tau}}
\def\dxi{\dot{\xi}}
\def\ddelta{\dot{\delta}}

\def\bsigma{\bar{\sigma}}
\def\bomega{\bar{\omega}}
\def\bpsi{\bar{\psi}}
\def\w{\omega}
\def\bal#1\eal{\begin{align}#1\end{align}}
\def\x{\xi}

\newcommand{\nn}{\nonumber}

\title{Higher spin holography with Galilean symmetry in general dimensions}
\author{Yang Lei $^{\clubsuit}$, and}
\author{Cheng Peng {$^\diamondsuit$} }

\affiliation[\clubsuit~~]{Centre for Particle Theory, Department of Mathematical Sciences, 
	Durham University, South Road, Durham DH1 3LE }
\affiliation[\diamondsuit~~]{Institut f\"ur Theoretische Physik, ETH Zurich, 	
CH-8093 Z\"urich, Switzerland }

\emailAdd{yang.lei@durham.ac.uk}
\emailAdd{pengch@itp.phys.ethz.ch}

\abstract{We construct Schr\"{o}dinger-like solutions of the Vasiliev higher spin theory in $D>3$ dimension. Symmetries of such solutions and the linearised equation of motion for the scalar on such backgrounds are analysed. We further propose Galilean invariant bosonic and fermionic field theories that could be dual to the two parity invariant higher spin theories on the Schr\"{o}dinger-like background respectively. The discussion is phrased mainly in $D=4$ dimension, while similar constructions follow straightforwardly in higher dimensions.}

\begin{document}

\maketitle

\section{Introduction}

Theories with higher spin symmetry have proved to be ideal playgrounds that help broaden and deepen our understanding of the holographic principle and String theory.  
The gauge/gravity duality with higher spin symmetry is a weak-weak duality; 
therefore, it is sensible to compare results from perturbative computations on both sides, which helps to understand the duality better \cite{Sezgin:2002rt, Klebanov:2002ja, Sezgin:2003pt, Koch:2010cy, Gaberdiel:2010pz, Douglas:2010rc,Gaberdiel:2011wb,Giombi:2011ya, Giombi:2011kc, Maldacena:2011jn,Maldacena:2012sf, Aharony:2012nh,Vasiliev:2012vf,Gaberdiel:2013vva,Giombi:2013fka, Candu:2013fta, Giombi:2014iua,Giombi:2014yra,Gaberdiel:2014yla,Creutzig:2014ula, Candu:2014yva, Hikida:2015nfa, Creutzig:2015hta}. 
In addition, remarkable progress has been made to discover the connection between higher spin theory and String theory \cite{Chang:2012kt,Gaberdiel:2014cha,Gaberdiel:2015mra,Baggio:2015jxa,Gaberdiel:2015uca,Bagchi:2015nca}, which supports the general belief that the higher spin theory should be identified as a subsector of String theory in the tensionless limit \cite{Gross:1987ar, Gross:1988ue, Witten:1988zd, Moore:1993qe, Sagnotti:2013bha, Sundborg:2000wp, Witten}.   

Previous examples of higher spin (super)gravity in $D \geq 3$ dimension are mostly on the background of maximally symmetric spacetime with non-zero cosmological constant, especially the Anti de-Sitter spacetime \cite{Vasiliev:1999ba,Vasiliev:2004cp,Bekaert:2005vh}.  In $D>3$ dimension, there are other classical solutions \cite{Sezgin:2005pv,Witten:1988zd, Didenko:2008va,Gubser:2014ysa} of the Vasiliev higher spin theory, but the geometric meanings of these solutions are less clear. 
While in $D=3$ dimension, solutions with different geometries \cite{Gary:2012ms, Afshar:2012nk} are found thanks to the Chern-Simons formulation of the higher spin theory \cite{Achucarro:1987vz,Witten:1988hc,Campoleoni:2010zq,Gaberdiel:2012uj}.  In particular, such solutions include spacetime with the 
Galilean isometry \cite{Son:2008ye, Balasubramanian:2008dm}, whose metric reads 
\begin{equation}\label{DSchroedinger}
ds^2 = -\frac{dt^2}{r^{2z}}+ \frac{dr^2 +2dt d\xi +\sum_{i=1}^{D-3}dx_i^2}{r^2}\,,
\end{equation}
where $D=3$ and $z>1$ is the dynamical scaling. 
Translation between the Chern-Simons formalism and the metric like formalism  
are discussed in \cite{Campoleoni:2010zq,Campoleoni:2011hg,Campoleoni:2012hp,Fujisawa:2012dk, Campoleoni:2014tfa, Lei:2015ika}. Further discussion of this  Schr\"{o}dinger geometry in holographic theories can be found in \cite{Guica:2010sw}. 

In general dimensions, the Schr\"{o}dinger spacetime \eqref{DSchroedinger} with $z\neq 2$ has the  Galilean group in $D-3$ spatial dimensions as isometry group. For $z=2$, the isometry group is enhanced to the  Schr\"{o}dinger group, which is the non-relativistic version of the conformal group \cite{Niederer:1972zz, Hagen:1972pd, Jackiw:1992fg,Nishida:2007pj, Blau:2009gd}. This symmetry governs many non-relativistic systems, such as unitary fermions, via the gauge/gravity duality \cite{Son:2008ye}. Properties of the Schr\"{o}dinger geometries have been investigated in different contexts \cite{Ross:2009ar,Guica:2010sw,Andrade:2014kba}, 
and several ways of embedding the  Schr\"{o}dinger geometry into String theory have been considered \cite{Herzog:2008wg, Maldacena:2008wh,Adams:2008wt}.

In this paper, we construct  Schr\"{o}dinger-like solutions,  whose metric has a form of  Schr\"{o}dinger geometry \eqref{DSchroedinger} with integer dynamical exponent $z$, to the Vasiliev equations in general $D>3$ dimension. This provides another interesting example of the higher spin holography in addition to the relatively well understood higher spin systems on the AdS background. The idea of our construction is, in the language of general relativity, to turn on higher spin fields which back-react on the geometry and support the non-maximally symmetric  Schr\"{o}dinger geometry. In practice, this amounts to solve the Vasiliev system in the ground state with non-vanishing (finite) higher spin fields.
This type of solution provides us a perfect example to examine  properties of the original Vasiliev system. For example, it demonstrates what is the precise effect of the higher spin fields on the geometry. In addition, it could make possible to extract more information of higher spin interactions by isolating some spin-s fields in the Vasiliev equation. Moreover, since most of the interesting condensed matter systems with Schr\"{o}dinger symmetry live in $d_s \geq 1$ spatial dimension, the corresponding bulk geometry should live in $D=d_s+3\geq 4$ dimension; this also motivates our construction.

Our paper is organized as follows. In section 2, we  review the Schr\"{o}dinger spacetime as a solution of $D=3$ dimensional  $hs[\lambda]$ higher spin theory in the Chern-Simons formulation. In section 3, we discuss how to find $D=4$ dimensional Schr\"{o}dinger-like solution  in the spinorial \cite{Giombi:2012ms, Vasiliev:1999ba} language. We show that the spacetime symmetry of the solution does not possess the whole Schr\"{o}dinger symmetry but only a subgroup of it. Therefore, the terminology ``Schr\"{o}dinger   
solution" simply refers to solutions  
whose corresponding metrics are of the form of \eqref{DSchroedinger}. In addition, linearised scalar equation of motion is analysed. 
In section 4, we briefly discuss the Schr\"{o}dinger geometry in general dimension in the vectorial language \cite{Vasiliev:2004cp, Bekaert:2005vh}. Field theories 
dual to these Schr\"{o}dinger solutions are proposed in section 5. 
We conclude our paper in section 6 and in particular we comment on one realization of $D=3$ dimensional Lifshitz higher spin theory with $z=2$ from dimensional reduction of Schr\"{o}dinger spacetime with $z=0$ in $D=4$ dimension \cite{Chemissany:2011mb, Chemissany:2012du}.

\section{Review of $3D$ Schr\"{o}dinger solution}

\subsection{Chern-Simons formulation}

We start with higher spin theory in the Chern-Simons formulation, which is defined as the difference of two Chern-Simons actions: 
\begin{equation}
S_{EH}= S_{CS}[A]- S_{CS}[\bar{A}]\,, \qquad S_{CS} = \frac{k}{4\pi}\int_M \text{Tr}(A \wedge dA+\frac{2}{3}A\wedge A\wedge A)\,,\label{Chernsimons}
\end{equation}
where $A^a$ and $\bar{A}^a$ take values in some Lie algebra. 

The equations of motion are the flatness conditions 
\begin{equation}\label{flatnesscondition}
F=dA+ A\wedge A=0\,, \qquad \bar{F}=d\bar{A}+\bar{A} \wedge \bar{A}=0\,.
\end{equation}
Metric-like fields in $3D$  
\eqref{DSchroedinger} are obtained by 
\begin{equation}\label{3Dmetricdefine}
g_{\mu\nu}= \frac{1}{2}\text{Tr}(E_\mu E_\nu)\,, \qquad \phi_{\mu\nu \rho} = \frac{1}{6} \text{Tr}(E_\mu E_\nu E_\rho)\,, \quad  \text{with} \quad E_\mu =\frac{1}{2}(A_\mu-\bar{A}_\mu) \,.
\end{equation}
A gauge transformation takes the connection to the form 
\bal 
A=e^{-\r L_0} (a(z)+\pa_z) e^{\r L_0}\,,\qquad \bar{A}=e^{\r L_0} (\bar{a}(\bar{z})-\pa_{\bar{z}}) e^{-\r L_0}\,.
\eal
Schr\"{o}dinger solutions with integer dynamical exponent $z$ \cite{Gary:2012ms, Lei:2015ika} \footnote{ We only focus on integer $z$ since it is recently argued \cite{Lei:2015ika} that Schr\"{o}dinger solution with fractional dynamical exponent $z$ in $3D$ higher spin theory may not have well-defined metric like description. The reason is that given the frame field $E_\mu$ defined above, one cannot solve spin-connection $\omega$ uniquely from torsion free equation \cite{Campoleoni:2012hp}
\begin{equation}
de + e\wedge \omega +\omega \wedge e =0\,.
\end{equation}
This implies that there is some information in the metric like fields that cannot be retrieved from Chern-Simons gauge fields. In addition,	we do not consider the Lifshitz spacetime \cite{Kachru:2008yh} 
\begin{equation}\label{Lifshitz}
ds^2 = -\frac{dt^2}{r^{2z}}+ \frac{dr^2+ dx_i^2}{r^2}\,,
\end{equation}
in higher spin context \cite{Gary:2014mca} with any $z$ for the same reason. }
can be constructed as
\begin{equation} \label{intS}
a = ( L_1 + \sigma W_+) dt\,, \quad \bar a = \sigma W_- dt + 2 L_{-1} d\xi \,,
\end{equation}
where $L_{0,\pm 1}$ are the modes of the Virasoro generator and the $W_{\pm}$ satisfy
\begin{equation}
[W_{\pm}, L_0] = \pm z W_{\pm}, \quad [W_{\pm}, L_{\pm 1}] = 0, \quad \text{tr}(W_+ W_-) \neq 0\,.
\end{equation}
We can simply take  $W_{\pm}$ to be the $V^{(z+1)}_{\pm(z)}$ modes in the higher spin algebra. 
In $z=2$ case, the spacetime metric and the spin-3 fields are 
\bal
& ds^2 = -\sigma^2 e^{4\rho}dt^2 +2e^{2\rho} dt d\xi +d\rho^2 \,,\\
& \phi_{t\xi\xi}=\frac{\sigma}{3}e^{4\rho}\,, \qquad \phi_{ttt}=-\frac{\sigma}{4} e^{4\rho} \,.\label{Schrhigherspin}
\eal
These metric like fields solve the Einstein equation perturbatively \cite{Lei:2015ika}.

\subsection{Vasiliev formulation }

Our normalization in this section is slightly different from \cite{Prokushkin:1998bq} but is self-consistent. Let us introduce oscillators $\hat{y}_\alpha \ (\alpha=1,2)$ fulfilling 
\begin{equation}
[\hat{y}_\alpha, \hat{y}_\beta ] = \frac{1}{2} \epsilon_{\alpha \beta} (1+\nu k)\;, \qquad  k \hat{y}_\alpha =-\hat{y}_\alpha k\;, \qquad k^2=1\,,
\end{equation}
where $\nu$ is a free parameter and $k$ is the Klein operator. 
Define bilinear oscillators $T_{\alpha \beta}$ 
\begin{equation}
T_{\alpha \beta}=\{\hat{y}_\alpha, \hat{y}_\beta\}\,,
\end{equation}
 that generate a $sl(2)$ algebra
\begin{eqnarray}
 &&  [T_{\alpha \beta}, T_{\gamma \sigma}] = \epsilon_{\beta \gamma} T_{\alpha \sigma} + T_{\beta \sigma} \epsilon_{\alpha \gamma} + T_{\alpha \gamma} \epsilon_{\beta \sigma} + \epsilon_{\alpha \sigma} T_{\beta \gamma}\,.
\end{eqnarray}
Higher (symmetric) powers of these oscillators give the higher spin generators. 
The connection with the Chern-Simons formulation is explained in section \ref{lonestar}.

In the current case, the gravitational connection  
\bal & W=\omega +\frac{1}{l} \psi e\,,\quad  
\psi^2=1\,, \quad [\psi, \hat{y}_\alpha]=0\,,\eal  where $\psi$ is the central involutive element and $l$ is the AdS radius, satisfies the  
equation of motion  \cite{Prokushkin:1998bq}
\bal \nn & dW +W\wedge  W=0\,. \quad
\eal The $z=2$ Schr\"{o}dinger gauge fields \eqref{intS} translate to the oscillator form 
\begin{eqnarray}
e &=& l(\frac{1}{4}r T_{11}+ \frac{\sigma}{8}r^2 T_{11}T_{11} -\frac{\sigma}{8}r^2 T_{22}T_{22})dt -\frac{l}{2}r T_{22}d\xi + \frac{l}{2r}T_{12}dr\,,\\
\omega &=& (\frac{1}{4}r T_{11} +\frac{\sigma}{8}r^2 T_{11}T_{11} +\frac{\sigma}{8}r^2 T_{22}T_{22})dt + \frac{1}{2}r T_{22}  d\xi\,.
\end{eqnarray}
via \eqref{sl2id}. It is then trivial to check that they solve the above equation of motion (setting $l=1$), which in component form reads
\begin{eqnarray}
\nn \text{Torsion free equations  } && \\ \label{torsionfree3d}
\psi\,T_{\alpha \beta}: && de^{\alpha \beta} + e^{\alpha \kappa}\wedge \omega^{\gamma \beta} \epsilon_{\kappa \gamma} + e^{\kappa \beta} \wedge \omega ^{\gamma \alpha} \epsilon_{\kappa \gamma} =0 \,,\\ \nonumber
\psi\,\sigma  T_{\alpha \beta} T_{\gamma \kappa}: && de^{\alpha \beta \gamma \kappa}+ 2\omega^{\alpha n \gamma \kappa} \wedge e^{m\beta} \epsilon_{nm} + 2 \omega^{\alpha \beta \gamma n} \wedge e^{m \kappa} \epsilon_{nm} \\ \label{spin33D1}
&&+ 2e^{\alpha n \gamma \kappa} \wedge \omega^{m\beta} \epsilon_{nm} + 2 e^{\alpha \beta \gamma n} \wedge \omega^{m \kappa} \epsilon_{nm}=0 \,,\\ \label{spin43D1}
\psi\, T_{\alpha \beta} T_{\gamma \kappa} T_{mn}: && \omega^{\alpha \beta \gamma \kappa} \wedge e^{mncd}=0\,,\\\label{AdS3D1}
\nn \text{ and Curvature equations  } && \\
T_{\alpha \beta}: && d\omega^{\alpha \beta} +\omega^{\alpha \kappa}\wedge \omega^{\gamma \beta} \epsilon_{\kappa \gamma} + \frac{1}{l^2} e^{\alpha \kappa} \wedge \omega ^{\gamma \beta} \epsilon_{\kappa \gamma} =0 \,,\\ \nonumber
\sigma T_{\alpha \beta} T_{\gamma \kappa}: && d\omega ^{\alpha \beta \gamma \kappa}+ 2\omega^{\alpha n \gamma \kappa} \wedge \omega^{m\beta} \epsilon_{nm} + 2 \omega^{\alpha \beta \gamma n} \wedge \omega^{m \kappa} \epsilon_{nm} \\ \label{spin33D2}
&&+ \frac{1}{l^2} (2e^{\alpha n \gamma \kappa} \wedge e^{m\beta} \epsilon_{nm} + 2 e^{\alpha \beta \gamma n} \wedge e^{m\kappa} \epsilon_{nm}) =0 \,,\\ \label{spin43D2}
T_{\alpha \beta} T_{\gamma \kappa} T_{mn}: && \omega^{(4)} \wedge \omega^{(4)} +e ^{(4)} \wedge e^ {(4)}=0\,. \label{AdS3D2}
\end{eqnarray}
This solution has no non-trivial holonomy, so one can do a large gauge transformation to relate this solution to empty AdS \cite{Lei:2015ika}.

\subsection{Scalar equations}
In this section, we consider the motion of a scalar in the above $3D$  Schr\"{o}dinger background, characterized by
\begin{equation}\label{KG3DHS}
dC +A*C -C* \bar{A} =0\,.
\end{equation}
We briefly review the analysis of \cite{Ammon:2011ua} in terms of the lone-star product in this subsection. The notation and its relation with the previously mentioned oscillator formalism is explained in Appendix \ref{lonestar}.

All the fields take value in the higher spin algebra
\begin{equation}
C= \sum_{s=1}^{\infty} \sum_{|m|<s} C_m^s V_m^s \,, \qquad 
A= \sum_{s=2}^{\infty} \sum_{|m|<s} A_m^s V_m^s\,, \qquad \bar{A} =\sum_{s=2}^{\infty} \sum_{|m|<s} \bar{A}_m^s V_m^s\,,
\end{equation}
with $C^1_0$ being the physical scalar. We now extract the equation of motion of $C_0^1$. If $A$ and $\bar{A}$ span pure $AdS_3$ gravity, equation \eqref{KG3DHS} reduces to Klein-Gordon equation. Now consider $z=2$ Schr\"{o}dinger spacetime \cite{Gary:2012ms, Afshar:2012nk}
\begin{eqnarray}
A&=& (\sigma e^{2\rho} V_2^3+ e^\rho V_1^2 )dt +V_0^2 d\rho\,, \quad  
\bar{A} = \sigma e^{2\rho} V_{-2}^3 dt +2e^{\rho} V_{-1}^2 d\xi -V_0^2 d\rho  \,,
\end{eqnarray}
where the constant source $\sigma$  parametrizes the higher spin deformation.  
Plugging these expansions into the scalar equation \eqref{KG3DHS} we get an infinite set of equations, one from each term proportional to $V^s_{m} dx^\mu\equiv V^s_{m,\mu}$. 
Remarkably, as shown in \cite{Ammon:2011ua}, we can choose a set of equations, being the coefficients of $\{V_{0,\rho}^1 ,V_{0,\bar{t}}^1, V_{1,x}^2, V_{0,\rho}^2, V_{0,x}^1, V_{-2,x}^3  , V_{-1,x}^2 , V_{-1,\rho}^2 , V_{-1,\rho}^3\}$,\footnotemark\footnotetext{Our choice is slightly different from that in \cite{Ammon:2011ua}.} that reduce to the explicit equation of motion for $C^1_0$
\begin{eqnarray}\nonumber
&& (\sigma e^{4\rho} \partial^4_\rho + 8\sigma e^{4\rho} \partial^3_\rho +2\sigma(11-\lambda^2) e^{4\rho} \partial^2_\rho -8\sigma e^{4\rho}(\lambda^2-3) \partial_\rho +\sigma e^{4\rho}(\lambda^2-1) (\lambda^2-9) \\ \label{Scalarequation3Dschr}
&& +2e^{2\rho} (1-\lambda^2)\partial_x +4e^{2\rho} \partial_\rho \partial_x +2e^{2\rho} \partial_x \partial^2_\rho- \sigma \partial^4_x +4 \partial_t \partial^2_x) C_0^1 =0\,.
\end{eqnarray}
Furthermore, as $\sigma \to 0$, one gets the $x$-derivative of the Klein-Gordon equation in AdS background \cite{Ammon:2011ua}; thus, we can solve the full equation perturbatively with respect to $\sigma$.

\section{$4D$ solution with Schr\"{o}dinger isometry}

\subsection{Star product in 4D}
Most of the notation in this section will follow \cite{Giombi:2012ms,Giombi:2009wh}, where  $x^\mu \ (\mu=0,1,2,3)$ denote spacetime Poincar{\'e} coordinates with $x_2=r$. In this coordinate, the AdS spacetime metric is \begin{equation}\label{AdSmetricPoincare}
ds^2= \frac{-dx_0^2 +dx_1^2 +dr^2+dx_3^2}{r^2}\,.
\end{equation}
The internal twistor space is parametrized by spinors $(Y,Z)=(y^{\alpha}, \bar{y}^{\dalpha},z^{\alpha}, \bar{z}^{\dalpha})$, $\alpha,\dot{\alpha}=1,2$. Here $z^{\alpha}, \bar{z}^{\dalpha}$ are auxiliary coordinates; physical fields are those with constraints $z^\alpha =\bar{z}^{\dalpha}=0$.

\noindent The star product of two spinor-valued functions can be defined as \cite{Giombi:2012ms}
\begin{eqnarray}\nonumber
f(Y,Z)* g(Y,Z) &=& f(Y,Z) \exp\Big[ \epsilon^{\alpha\beta} (\overleftarrow{\partial}_{y^{\alpha}}+ \overleftarrow{\partial}_{z^{\alpha}})(\overrightarrow{\partial}_{y^{\beta}}- \overrightarrow{\partial}_{z^{\beta}}) \\
&&+ \epsilon^{\dalpha\dbeta} (\overleftarrow{\partial}_ {\bar{y}^{\dalpha}}+ \overleftarrow{\partial}_ {\bar{z}^{\dalpha}})(\overrightarrow{\partial}_{\bar{y}^{\dbeta}}- \overrightarrow{\partial}_{ \bar{z}^{\dbeta}}) \Big] g(Y,Z)\,.
\end{eqnarray}
There are in addition Klein operators 
$K(t)= e^{t z^{\alpha} y_\alpha}$ and $\bar{K}(t)= e^{t \bar{z}^{\dalpha} \bar{y}_{\dalpha}}$.

Vasiliev master fields include a gravitational connection $W= W_\mu(x|y,\bar{y},z,\bar{z})dx^\mu$, an auxiliary fields $S= dz^\alpha S_\alpha (x|y,\bar{y},z,\bar{z}) +d\bar{z}^{\dalpha} S_{\dalpha} (x|y,\bar{y},z,\bar{z}) $ \footnote{The spinor indices are raised and lowered by the antisymmetric tensor $\epsilon_{\alpha \beta}$, \bal
\nn	A^\alpha =\epsilon^{\alpha \beta} A_\beta; \quad A_\alpha =A^\beta \epsilon_{\beta \alpha}, \quad \epsilon_{12} =\epsilon^{12} =1\,.
	\eal } and a scalar field $B(x|y,\bar{y},z,\bar{z})$.  
The equations of motion that determine the dynamics of the system are
\begin{subequations}\label{4Dflatequation}
\begin{align} 
& d_x W +W*\wedge W=0\,, \label{wweqn} \\
& d_Z W +d_xS +\{ W,S\}_* =0 \,,\label{WSequation}\\
& d_Z S +S*  S =B*K dz^2 + B* \bar{K} d\bar{z}^2 \,,\\
&  d_x B+ W*B-B*\pi(W)=0\,,  \label{last4Dequation}\\
& d_Z B+ S*B -B*\pi(S)=0\,, 
 \end{align}
\end{subequations}
where $\pi(H)$ flips the signs of unbarred spinors $(y,z,dz)$ in $H$ while it preserves the signs of barred coordinates $(\bar{y},\bar{z},d\bar{z})$. These master fields also satisfy 
\begin{equation}
[R,W]_* =\{ R,S\}_* =[R,B]_* =0\,,
\end{equation}
where $R=K \bar{K}$. This implies $W,B$ are even functions of $(Y,Z)$ while $S$ is an odd function of $(Y,Z)$.

In this section, we will discuss the vacuum solutions of master equation \eqref{4Dflatequation}, i.e. $B=0\,,~ S=dz^{\alpha}\,z_{\alpha}+d\bar{z}^{\dot{\alpha}}\,\bar{z}_{\dot{\alpha}}$ and $W(Y,Z)=W(Y)$ from \eqref{WSequation}.

\subsection{AdS solution in lightcone coordinate}

Vacuum $AdS_4$ spacetime 
\bal
B=0\,,\quad S=dz^{\alpha}\,z_{\alpha}+d\bar{z}^{\dot{\alpha}}\,\bar{z}_{\dot{\alpha}}\,,\quad W=e_{\alpha \dbeta} y^{\alpha} \bar{y}^{\dbeta}+\omega_{\alpha \beta} y^{\alpha} y^{\beta} +\omega_{\dalpha \dbeta} \bar{y}^{\dalpha} \bar{y}^{\dbeta}\,,\label{3dsol}
\eal
is a solution to the Vasiliev equations \eqref{4Dflatequation}, which reduces to the component form 
\begin{eqnarray}\label{311}
y^{\alpha} \bar{y}^{\dalpha}: && de_{\alpha \dalpha} + 4e_{\gamma \dalpha } \wedge \omega_{\alpha \beta} \epsilon^{\gamma \beta}- 4e_{\alpha \dbeta} \wedge \omega_{\dalpha \dgamma} \epsilon^{\dgamma \dbeta}=0 \,,\\
y^{\alpha} y^{\beta}: && d\omega_{\alpha \beta}+ e_{\alpha \dgamma} \wedge e_{\beta \dkappa} \epsilon^{\dgamma \dkappa} +4 \omega_{\beta \kappa} \wedge \omega_{\alpha \gamma} \epsilon^{\kappa \gamma}=0  \,,\\ \label{313}
\bar{y}^{\dalpha} \bar{y}^{\dbeta}: && d\omega_{\dalpha \dbeta} - e_{\kappa \dalpha } \wedge e_{\gamma \dbeta} \epsilon^{\gamma \kappa} +4 \omega_{\dbeta \dkappa} \wedge \omega_{\dalpha \dgamma} \epsilon^{\dkappa \dgamma}=0\,.
\end{eqnarray}
Explicitly, we have
\begin{equation}
e_{\alpha \dbeta} = \tfrac{1}{4}e^a (\sigma_a)_{\alpha \dbeta}\,, \quad \omega_{\alpha \beta} = -\omega^a (\sigma_{a2} \epsilon)_{\alpha \beta}\,, \quad \bomega_{\dalpha \dbeta}=-\omega^a (\epsilon \bsigma_{a2})_{\dalpha \dbeta}\,,\label{3dew}
\end{equation}
where $e^a=\tfrac{\delta^a_\mu}{r}dx^\mu\,,~ \omega^a =\tfrac{\delta^a_\mu}{8r}dx^\mu $ are the veilbein and the spin connection of AdS spacetime \eqref{AdSmetricPoincare} in the lightcone Poincar{\'e}  coordinate 
\begin{equation}\label{lightconeAdS4}
ds^2 = \frac{2 dt d\xi+dr^2+dx^2}{r^2}\,,\qquad \xi= \frac{x_1-x_0}{\sqrt{2}}\,, ~ t=\frac{x_1+x_0}{\sqrt{2}}\,,~ x=x_3\,.
\end{equation}
We have further employed Pauli matrices in the lightcone coordinate in  \eqref{3dew}
\begin{eqnarray}
\nn && \sigma_t = \tfrac{\sigma_0 +\sigma_1}{\sqrt{2}}, \qquad \sigma_\xi = \tfrac{-\sigma_0 +\sigma_1}{\sqrt{2}}, \qquad
\sigma_r =\sigma_2, \qquad \sigma_x =\sigma_3 \,,\\
&& \sigma_{t\mu} = \tfrac{\sigma_{0\mu}+\sigma_{1\mu}}{\sqrt{2}}, \quad \sigma_{\xi\mu} = \tfrac{-\sigma_{0\mu}+\sigma_{1\mu}}{\sqrt{2}}, \quad
\bsigma_{t\mu} = \tfrac{\bsigma_{0\mu} +\bsigma_{1\mu}}{\sqrt{2}}, \quad \bsigma_{\xi\mu} = \tfrac{-\bsigma_{0\mu} +\bsigma_{1\mu}}{\sqrt{2}}\,.
\end{eqnarray}
Further notice that we work in the Minkowski signature, so the Pauli matrices are the familiar ones that are hermitian. As a consequence, the parity action is our convention is then 
\bal 
y_\alpha \leftrightarrow \bar{y}_{\dot{\alpha}}\,, \quad z_\alpha \leftrightarrow \bar{z}_{\dot{\alpha}}\,,   
\eal
and further accompanied with hermitian conjugation of the coefficients of the oscillators.

\subsection{Schr\"{o}dinger solution with $z=2$}

We are now ready to construct $4D$ Schr\"{o}dinger geometry \eqref{DSchroedinger} in Vasiliev higher spin theory. The simplest non-trivial example is the $z=2$ Schr\"{o}dinger geometry which turns out to be supported by extra $s=3$ higher spin fields. 
We consider a variant form of the Schr\"{o}dinger metric  \begin{equation}\label{schmetric}
ds^2 = -\frac{\sigma^2 dt^2}{r^{2z}}+ \frac{2dtd\xi +dr^2 +dx^2}{r^2}\,,\qquad z=2\,, \quad \sigma\in \mathbb{R}\,, \sigma\neq 0\,, 
\end{equation}
which can be converted from \eqref{DSchroedinger}  by field redefinition $t \to \sigma t, \ \xi \to \xi \sigma^{-1}$. 

\subsubsection{General solution}
We try to find a ground state solution to 
\eqref{4Dflatequation} of the form
\bal
B=0\,,\qquad S= dz^{\alpha}\,z_{\alpha}+d\bar{z}^{\dot{\alpha}}\,\bar{z}_{\dot{\alpha}}\,,\qquad W(Y,Z|x)=W(Y|x)\,,
\eal
with some spin-3 fields turned on in $W$. We simply take $W= W_2 +W_3$, where $W_2$ is the spin-2 piece \eqref{3dsol}, \eqref{3dew}, and $W_3$ encodes spin-3 fields that are quartic in the $y,\bar{y}$ oscillators
{\small\bal
W_3 = \omega_{\alpha \beta \gamma \kappa} y^\alpha y^\beta y^\gamma y^\kappa + \omega_{\alpha \beta \gamma \dkappa} y^\alpha y^\beta y^\gamma \bar{y}^{\dkappa} + \omega_{\alpha \beta \dgamma \dkappa} y^\alpha y^\beta \bar{y}^{\dgamma} \bar{y}^{\dkappa} +\omega_{\alpha \dbeta \dgamma \dkappa} y^\alpha \bar{y}^{\dbeta} \bar{y}^{\dgamma} \bar{y}^{\dkappa} + \omega_{\dalpha \dbeta \dgamma \dkappa}  \bar{y}^{\dalpha} \bar{y}^{\dbeta} \bar{y}^{\dgamma} \bar{y}^{\dkappa}\,.\label{generalspin3ansatz}
\eal}
The only nontrivial equation \eqref{wweqn} decomposes schematically to
\begin{subequations} \label{AdS4Dequation}
	\bal
y^2:  & \quad d_x W_2 + W_2 *\wedge W_2 =0\,,  \label{4DSHCRmaster}\\
y^4:  & \quad d_x W_3 +W_2 *\wedge W_3 +W_3 *\wedge W_2 =0 \,,\label{4DSHCRmaster2}\\
y^6:  &  \quad W_3 * \wedge W_3 =0\,. \label{4DSHCRmaster3}
\eal
\end{subequations}
The equation \eqref{4DSHCRmaster} simply means we can take $W_2$ as the AdS connection \eqref{3dsol} and \eqref{3dew}. The equation \eqref{4DSHCRmaster3} is very restrictive and can only be solved due to the wedge product: we take $W_3$ to be proportional to $dt$ in the light of our aimed solution \eqref{DSchroedinger}. 
The only remaining equation to be solved, namely \eqref{4DSHCRmaster2}, decomposes to
{\small
\bal
\nn y^4: &~~ d\omega_{\alpha \beta \gamma \kappa} + 2e_{\alpha \dxi} \wedge \omega_{\beta \gamma \kappa \ddelta} \epsilon^{\dxi \ddelta} +16 \omega_{\alpha \xi} \wedge \omega_{\beta \gamma \kappa \delta} \epsilon^{\xi \delta} =0\,, \\ \nonumber
y^3 \bar{y}: &~~ d\omega_{\alpha \beta \gamma \dkappa} + 8 e_{\xi \dkappa} \wedge \omega_{\alpha \beta \gamma \delta} \epsilon^{\xi \delta} + 4 e_{\alpha \dxi} \wedge \omega_{\beta \gamma \ddelta \dkappa} \epsilon^{\dxi \ddelta}+ 12\omega_{\alpha \xi} \wedge \omega_{\beta \gamma \delta \dkappa} \epsilon^{\xi \delta}  + 4 \omega_{\dkappa \dxi} \wedge \omega_{\alpha \beta \gamma \ddelta} \epsilon^{\dxi \ddelta} =0 \,,\\ \nonumber
y^2 \bar{y}^2: &~~ d\omega_{\alpha \beta \dgamma \dkappa} + 6 e_{\xi \dgamma} \wedge \omega_{\alpha \beta \delta \dkappa} \epsilon^{\xi \delta} + 6 e_{\alpha \dxi} \wedge \omega_{\beta \dgamma \dkappa \ddelta} \epsilon^{\dxi \ddelta} +8 \omega_{\alpha \xi} \wedge \omega_{\beta \delta \dgamma \dkappa} \epsilon^{\xi \delta} +8 \omega_{\dgamma \dxi} \wedge \omega_{\alpha \beta \dkappa \ddelta} \epsilon^{\dxi \ddelta}=0 \,,\\ \nonumber
y \bar{y}^3: &~~ d\omega_{\alpha \dbeta \dgamma \dkappa} + 4 e_{\xi \dbeta} \wedge \omega_{\alpha \delta \dgamma \dkappa} \epsilon^{\xi \delta} + 8e_{\alpha \dxi} \wedge \omega_{\dbeta \dgamma \dkappa \ddelta} \epsilon^{\dxi \ddelta} + 4\omega_{\alpha \xi} \wedge \omega_{\delta \dbeta \dgamma \dkappa} \epsilon^{\xi \delta} +12 \omega_{\dbeta \dxi} \wedge \omega_{\alpha \dgamma \dkappa \ddelta} \epsilon^{\dxi \ddelta}=0\,,
\\ \nn
\bar{y}^4: &~~ d\omega_{\dalpha \dbeta \dgamma \dkappa} +2 e_{\xi \dalpha} \wedge \omega_{\delta \dbeta \dgamma \dkappa } \epsilon^{\xi \delta} +16 \omega_{\dalpha \dxi} \wedge \omega_{ \dbeta \dgamma \dkappa \ddelta} \epsilon^{\dxi \ddelta} =0\,.
\eal}
Considering only time independent, spherical symmetric solution, this set of equations is solved to get
\bal
& \nn \w_{2 {2} {2} {2}} =\tfrac{ C_1}{4 r^2}\,, \qquad  \w_{2 {2} {2}\dot{2}} =\tfrac{-i C_1}{r^2}\,, \qquad \w_{2 {2}\dot{2}\dot{2}} =\tfrac{-3  C_1}{2 r^2}\,,  \qquad  \w_{2\dot{2}\dot{2}\dot{2}} =\tfrac{iC_1}{r^2}\,, \quad \w_{\dot{2}\dot{2}\dot{2}\dot{2}} =\tfrac{ C_1}{4 r^2}\,,\\ 
\nn & \w_{2 {2} {2} {1}} =\tfrac{- C_2}{6 r^2}\,, \qquad  \w_{2 {2} {2}\dot{1}} =\tfrac{2 i C_2}{3 r^2}\,, \qquad \w_{2 {2}\dot{2}1} =\tfrac{2 i C_2}{3 r^2}\,, \qquad  \w_{2 {2}\dot{2}\dot{1}} =\tfrac{C_2}{r^2}\,, \\
& \nn \w_{2 \dot{2} \dot{2} {1}} =\tfrac{C_2}{r^2}\,, \qquad  \w_{2 \dot{2} \dot{2}\dot{1}} =\tfrac{-2 i C_2}{3 r^2}\,, \qquad \w_{\dot{2} \dot{2}\dot{2} 1} =\tfrac{-2 i C_2}{3 r^2}\,, \qquad  \w_{\dot{2} \dot{2}\dot{2}\dot{1}} =\tfrac{-C_2}{6 r^2}\,, \\
&  \nn \w_{1 {1} {1} {1}} =\tfrac{ C_3}{4 r^2}\,, \qquad  \w_{1 {1} {1}\dot{1}} =\tfrac{-iC_3}{r^2}\,, \qquad \w_{1 {1}\dot{1}\dot{1}} =\tfrac{-3  C_3}{2 r^2}\,, \qquad  \w_{1\dot{1}\dot{1}\dot{1}} =\tfrac{iC_3}{r^2}\,, \quad \w_{\dot{1}\dot{1}\dot{1}\dot{1}} =\tfrac{ C_3}{4 r^2}\,,\\
\nn & \w_{1 {1} {2} {2}} =\tfrac{- C_4}{6 r^2}\,, \qquad  \w_{1 {1} {2}\dot{2}} =\tfrac{2 i C_4}{3 r^2}\,, \qquad  \w_{1 {1}\dot{2}\dot{2}} =\tfrac{C_4}{r^2}\,, \qquad \w_{1 \dot{1}{2}2} =\tfrac{2 i C_4}{3 r^2}\,,  \qquad  \w_{1 \dot{1}{2}\dot{2}} =\tfrac{C_4}{r^2}\,,\\
\nn & \w_{\dot{1} \dot{1} {2} {2}} =\tfrac{C_4}{r^2}\,, \qquad  \w_{1 \dot{1} \dot{2}\dot{2}} =\tfrac{-2 i C_4}{3 r^2}\,, \qquad \w_{\dot{1} \dot{1}\dot{2} 2} =\tfrac{-2 i C_4}{3 r^2}\,, \qquad  \w_{\dot{1} \dot{1}\dot{2}\dot{2}} =\tfrac{-C_4}{6 r^2}\,,\\
\nn & \w_{1 {1} {1} {2}} =\tfrac{- C_5}{6 r^2}\,, \qquad  \w_{1 {1} {1}\dot{2}} =\tfrac{2 i C_5}{3 r^2}\,, \qquad \w_{1 {1}\dot{1}2} =\tfrac{2 i C_5}{3 r^2}\,, \qquad  \w_{1 {1}\dot{1}\dot{2}} =\tfrac{C_5}{r^2}\,, \\
&  \w_{1 \dot{1} \dot{1} {2}} =\tfrac{C_5}{r^2}\,, \qquad  \w_{1 \dot{1} \dot{1}\dot{2}} =\tfrac{-2 i C_5}{3 r^2}\,, \qquad \w_{\dot{1} \dot{1}\dot{1} 2} =\tfrac{-2 i C_5}{3 r^2}\,, \qquad  \w_{\dot{1} \dot{1}\dot{1}\dot{2}} =\tfrac{-C_5}{6 r^2}\,,  \label{z=2Schrsolution4D}
\eal
where $C_i \ (i=1,...,5)$ are arbitrary real constants. Furthermore, this solution is manifestly parity invariant.

We would like to remark that, in general, once spin-3 generators in $D>3$ dimensional higher spin theory is included, one is forced to include the infinite tower of higher spin fields to solve the equation. This problem is avoided in our construction since the spin-3 fields are only turned on in the $t$ direction and $dt \wedge dt=0$. For this reason we are able to isolate a single spin-s field, which back-reacts and supports the $z=s-1$ Schr\"{o}dinger spacetime. The spinorial index structure of $\omega_{(4)}$ fields implies that the above solution can be
expanded in a basis consisting of tensors of two Pauli matrices. 
Making use of the identity \cite{Wess:1992cp}
\begin{equation}
\sigma^\mu _{\alpha \dgamma} \sigma^\nu _{\beta \dkappa}  + \sigma^\nu _{\alpha \dgamma} \sigma^\mu _{\beta \dkappa} =  \eta^{\mu\nu} \sigma^r_{\alpha \beta} \sigma^r_{\dgamma \dkappa} +4 (\sigma^{\l \mu} \epsilon)_{\alpha \beta}  (\epsilon \bar{\sigma}^{l \nu})_{\dalpha \dbeta},
\end{equation}
the $W_3$ field can be recast into 
\begin{equation}
W_3=\big(e^{ab}\sigma_a \sigma_b+ H_{{\text{ew}}}^{ab}\, \sigma_a (\sigma_{b2} \epsilon)+H_{\overline{\text{ew}}}^{ab}\, \sigma_a  ( \epsilon \bar{\sigma}_{b2})
+H_{{\text{ww}}}^{ab}\, (\sigma_{a2} \epsilon) (\sigma_{b2} \epsilon)+H_{\overline{\text{ww}}}^{ab}\, ( \epsilon \bar{\sigma}_{a2})  ( \epsilon \bar{\sigma}_{b2})\big) dt\,.\label{schexp}
\end{equation} 
We have checked that the $e^{ab}, H^{ab}$ fields can be determined for the Schr\"{o}dinger spacetime \eqref{z=2Schrsolution4D}. However, the result is not much simpler than \eqref{z=2Schrsolution4D} and is not very illuminating so we do not show them explicitly.

Another comment is that given a generalised vielbein 
\bal E= e_{\alpha \dbeta} y^{\alpha} \bar{y}^{\dbeta} + \omega_{\alpha \beta \dgamma \dkappa} y^\alpha y^\beta \bar{y}^{\dgamma} \bar{y}^{\dkappa}\,,\eal which means fixing the $C_i$, $i=1,\ldots,5$ parameters, the $W$ field is fully determined. This is equivalent to the statement that (generalised) spin-connection can be fully determined by the (generalised) veilbein from ``torsion free" equations. Therefore, our $z=2$ Schr\"{o}dinger solution is free from degeneracy problem \cite{Lei:2015ika}. 

\subsubsection{The metric}
As we have briefly explained in the previous section, we do not treat the spin-3 fields as probe but take their backreaction on the geometry into account. 
We thus propose the following formula to compute the metric from the (generalised) vielbein 
\bal 
g= \text{Tr}(E*E)\,, \quad \label{4dgprop}
\eal
where the trace is defined in \eqref{trace}. Notice that this definition reduces to the more familiar definition $g=\text{Tr}(e*e)$ in general relativity when the higher spin fields are turned off.

This formula is determined by requiring the invariance of the metric under generalised local Lorentz transformations that rotate the local Lorentz indices and thus the local basis.  This idea was first proposed in 3-dimensional \cite{Campoleoni:2010zq} and we simply generalise it to higher dimension. To justify our proposal, we start with the general gauge transformation of any solution of the set of Vasiliev equations \eqref{4Dflatequation}
\begin{equation}
\delta W= d\epsilon +[W\,,\epsilon]_*\,, \quad \delta B=B*\pi(\epsilon)-\epsilon* B\,,\quad \delta S=[S,\epsilon]_* \,. \label{rigid}
\end{equation}
Since we have $B=0$ and $\epsilon=\epsilon(Y|x)$, we only consider the first transformation. From which we can read off the general transformation $\delta E$ of our definition $E$ \eqref{4dgprop}. Then we want to decompose the gauge transformation as
\bal
\epsilon=\xi+\Lambda+\Lambda_{\text{extra}}\,,
\eal 
where $\xi$ parametrizes the generalised diffeomorphisms, $\Lambda$ parametrizing the generalized local Lorentz transformations and $\Lambda_{\text{extra}}$ parametrizes the extra gauge transformation associated to the 
extra auxiliary fields. The difference between the latter two is that the $\Lambda$ only rotates the index in the first row in the two-row Young tableaux notation while $\Lambda_{\text{extra}}$ rotates indices in both the two rows. We thus require the metric to be invariant under all transformations parametrized by $\Lambda$.\footnote{The metric does transform under $\Lambda_{\text{extra}}$, which is the higher dimensional analogue of phenomena discussed in, e.g. \cite{Ammon:2011nk, Castro:2011fm}.} It can be explicitly checked that our proposal \eqref{4dgprop} fullfills this requirement: the extra variation of the vielbeins under the local higher spin transformation is cancelled by the variation of the generalised vielbeins $\omega_{\alpha \beta \dgamma \dkappa}$. 
In fact, there is a much easier way to demonstrate this invariance. The variation takes a nice form $\delta E =[E,\Lambda]_*$, 
then it is trivial to verify the invariance of the metric by cyclicity of the trace \footnote{We thank Stefen Theisen to point this out to us.}
\begin{equation}
\delta_\Lambda g =\text{Tr}([E,\Lambda]_* * E + E* [E,\Lambda]_*) =0\,.
\end{equation}
With this definition, the solution we have found gives the following metric
\bal
ds^2 = -(72C_4^2-64C_2C_5+144C_1C_3)\frac{ dt^2}{r^{4}}+ \frac{2dtd\xi +dr^2 +dx^2}{r^2}\,.\label{metric1}
\eal

\subsubsection{Higher spin fields}

The spin-3 metric like field can be determined similarly 
\bal
\Phi= \text{Tr}(E*E*E)\,,\label{defspin3}
\eal
which is again invariant under the higher spin generalisation of the local Lorentz transformation.  
Linearizing the above spin-3 field leads to traceless symmetric tensor
\begin{equation}
\Phi_{\mu\nu_1 \nu_2} \sim \text{Tr}(e_{\alpha_1 \dbeta_1} y^{\alpha_1} \bar{y}^{\dbeta_1} * e_{\gamma_1 \dkappa_1} y^{\gamma_1} \bar{y}^{\dkappa_1} * \omega_{\alpha_2 \beta_2 \dgamma_2 \dkappa_2} y^{\alpha_2} y^{\beta_2} \bar{y}^{\dgamma_2} \bar{y}^{\dkappa_"}) \sim \sigma^{\alpha_1 \dgamma_2}_{\nu_1} \sigma^{\beta_2 \dkappa_2}_{\nu_2} \omega_{\mu|\alpha_2 \beta_2 \dgamma_2 \dkappa_2},\label{spin3linear}
\end{equation}
which agrees with the expression given in \cite{Giombi:2012ms} up to normalization.
We can further evaluate the fully nonlinear spin-3 fields \eqref{defspin3} explicitly
\bal
\nn &\phi_{ttt}=\tfrac{3 \left((3  C_1+8 C_2+3  C_3-12 C_4-8 C_5) r^2+512 \left(4 C_4^3-9 C_1 C_3 C_4-8 C_2 C_5 C_4-6  C_1 C_5^2-6  C_2^2 C_3\right)\right)}{2 r^6}\,,\\
\nn & \phi_{tt\xi}=\tfrac{-4 C_4-3  (C_1+C_3)}{ r^4}\,, \quad  \phi_{t\xi\xi}=-\tfrac{-3  C_1+8 C_2-3  C_3+12 C_4+8 C_5}{2 r^4}\,,\quad \phi_{txx}=\tfrac{(3C_1+3C_3+4C_4)}{r^4}\,,\\
&\phi_{ttx}=\tfrac{\sqrt{2}(3  C_1+4 C_2-3  C_3-4 C_5)}{  r^4}\,,\qquad \phi_{tx\xi}= -\tfrac{3 C_1-4 C_2-3 C_3+4 C_5}{ \sqrt{2} r^4}\,,
\eal
with all other components vanish.
Notice that in most of the terms the power at the boundary is exactly the dimension $\Delta=4$ of a conserved spin-3 currents in the dual field theory. The only exception is the $r^{-6}$ term in $\Phi_{ttt}$ which has cubic coefficients $C_iC_j C_k$; both its scaling behavior and its coefficient structure indicate the non-linear nature of this term. We will discuss more about this $r^{-6}$ power in section \ref{fty}.

As we have shown explicitly, 
the metric and the spin-3 metric like fields can be uniquely determined. 
To determine metric like higher spin fields with $s>3$, more information is needed, which is similar to what happens in 3D \cite{Campoleoni:2011hg}, in addition to the requirement of local Lorentz invariance and the correct linearisation limit. This is because there are more than one combinations of veilbeins satisfying the above constraints. For example, for $s=4$, $(\text{tr}(E*E))^2$ and $\text{tr}(E*E*E*E)$ are both local Lorentz invariant. Only a linear combination of these two terms gives the right Fronsdal field 
\bal
\nn \Phi^{(4)} =\text{tr}\big((E*E*E*E)_s\big) + c~\text{tr}(E*E)\,\text{tr}(E*E)\,,
\eal 
where $(a*b)_s=a*b+b*a$ is the totally symmetric star product. 
The coefficient $c$ can be fixed by imposing the double-traceless condition or by imposing a Fefferman-Graham-like gauge condition $ \Phi_{rrrr}=0 $ \cite{Li:2015osa}. Remarkably, the two conditions lead to the same value $c= -\frac{1}{2}$.\footnote{The exact value of $c$ depends on our definition of the trace, but the conclusion that the two conditions lead to the same value is independent of our definition of the trace; the latter can be checked explicitly.}  This result agrees with our expectation and also agrees with what happens in 3D.

We comment here that even though we only turn on spin-2 and spin-3 components of the frame like field $W$ \eqref{generalspin3ansatz}, there can be a nonzero spin-4 metric like field as constructed above. This property can only be seen at the fully nonlinear level; the linearised spin-4 field, defined similarly as \eqref{spin3linear}, vanishes.  Moreover, we believe the whole tower of the metric like fields of arbitrary spin are nonzero unless protected by some hidden symmetries.

\subsubsection{Symmetries of the solution}\label{sec:sym}

\paragraph{Relation with the AdS spacetime}
One immediate question is if the solution we have got is gauge equivalent to the AdS vacuum. This is not true in the presence of boundary since the boundary behaviors of the two solutions, for instance the fall-off behaviors of the metric, are different and thus cannot be related by any gauge symmetry.  This statement actually does not depend on our definition of the metric and spin-3 fields; it can be concluded already from the master field $W$. As explicitly shown in the solution, we have \bal W^{\text{Sch}}=W^{\text{AdS}}+W^3\,, \eal where the $W^3$ contains the spin-3 generators and the boundary behavior is \bal
W^{\text{AdS}}\sim\tfrac{1}{r}\,,\qquad W^3\sim\tfrac{1}{r^2}\,.\eal Thus the boundary behavior of the two solutions are different and thus cannot be related by gauge transformation.

\paragraph{Spacetime symmetry}

We can find the spacetime symmetry of the full solution by finding all the killing vectors of both the metric and the higher spin metric like fields. By definition, the Lie derivative of the fields along the direction of any killing vector $\chi^\mu$ vanishes
\bal 
\mathcal{L}_\chi g_{\mu\nu}=0\,,\qquad  \mathcal{L}_{\chi} \phi_{\mu\nu\rho}=0\,,\qquad  \mathcal{L}_{\chi} \phi_{\mu\nu\rho\sigma}=0 ~\ldots \,.\label{lieder}
\eal
Solving the first equations, we find the follow killing vectors generating the Schroedingere isometry of the spacetime in our $z=2$ example
\bal 
&\chi_{H}=\partial_t\,,\qquad \chi_{M}=\partial_\xi\,,\qquad \chi_{P}=\partial_x\,, \qquad \chi_K=x\partial_\xi-t \partial_x\,,\\
& \chi_D=2t\partial_t+x\partial_x+r\partial_r\,,\qquad \chi_C=t^2 \partial_t-\tfrac{1}{2}(x^2+r^2)\partial_\xi+t x\partial_x +t r \partial_r \,.
\eal
Applying the Lie derivatives associated with these vectors to the spin-3 fields, we find in general only  $H,M,P$ remain symmetry of the spin-3 fields. However, for special choice of the parameters $C_i$, $i=1,\ldots,5$, the symmetry of the system could get enhanced. These extra enhanced symmetries can be summarized in  Table \ref{Table1}
	\begin{table}
	\renewcommand{\arraystretch}{1.2}
	\newcommand{\tabincell}[2]{\begin{tabular}{@{}#1@{}}#2\end{tabular}}
	\centering
	\caption{Symmetry enhancement and metric like fields}
	\label{Table1}
	\begin{tabular}{|c|c|c|c|}
		\hline
		 & Killing vectors & $-g_{tt}r^4$ & spin-3 fields \\
		\hline 
		(a) & $\chi_K$ & $162 C_3^2$ & $\phi_{ttt}= \tfrac{72  C_3}{r^4} $\\ 
		\hline 
		\rule{-3pt}{3ex}
		(b) & $\chi_D$ & $162 C_3^2$ & $\phi_{ttt}= -\tfrac{3072  C_3^3}{r^6}\,, ~\phi_{tt\xi}= \tfrac{-8  C_3}{r^4} \,, ~\phi_{txx}= \tfrac{ 8 C_3}{r^4} $\\ 
		\hline 
		\rule{-3pt}{5ex}
		(c) & $\chi_D-\sqrt{2}\chi_K $ & $\tfrac{9}{2} C_3^2$ & \tabincell{l}{ $\phi_{ttt}= \tfrac{6  \left(3 C_3 r^2-8 C_3^3\right)}{ r^6}\,, ~\phi_{tt\xi}= \tfrac{-2  C_3}{r^4}$\\ $\phi_{ttx}=\tfrac{6\sqrt{2}  C_3}{ r^4} \,, ~\phi_{txx}= \tfrac{2 C_3}{ r^4}$} \\ 
		\hline
		\rule{0pt}{8ex} 
		(d) & \tabincell{c}{{\footnotesize $\chi_D+\tfrac{2 \left(C_1 \mp\sqrt{C_1} \sqrt{C_3} \right)\chi_K}{\sqrt{2} C_1+\sqrt{2} \sqrt{C_3} \sqrt{C_1}} $}}  & \tabincell{c}{\tiny $\tfrac{9}{2} (C_1^2+34 C_3C_1 +C_3^2) $ } & \tabincell{l}{
			$\phi_{ttt}= \tfrac{6  \left(3 \left(\pm\sqrt{C_1}-\sqrt{C_3}\right)^2 r^2-8 \left(\pm\sqrt{C_1}+\sqrt{C_3}\right)^6\right)}{ r^6}\,$ \\ $\phi_{tt\xi}= \tfrac{-2  \left(\pm\sqrt{C_1}+\sqrt{C_3}\right)^2}{r^4},~\phi_{ttx}=\tfrac{6 \sqrt{2}  (C_1-C_3)}{r^4} $ \\
			$\phi_{txx}= \tfrac{2  \left(\pm\sqrt{C_1}+\sqrt{C_3}\right)^2}{r^4}\,.$ }\\\hline
	\end{tabular} 
	\renewcommand{\arraystretch}{1}
\end{table}
where the coefficients take the following values in different cases:
{\small\begin{eqnarray}
(a) :&&  C_2\to \tfrac{3}{2}  C_3, \quad C_1\to C_3, \quad  C_4 \to -\tfrac{3}{2}  C_3,\quad  C_5\to \tfrac{3}{2}   C_3 \,,\\ \nonumber
(b) :&&  C_2\to 0, \quad  C_1\to C_3, \quad C_4\to \tfrac{1}{2}  C_3,\quad  C_5\to 0\,,\\ \nonumber
(c) :&&  C_1\to 0, \quad  C_2\to 0, \quad  C_4\to -\tfrac{1}{4}  C_3,\quad  C_5\to \tfrac{3}{4}   C_3 \,,\\ \nonumber
(d) :&&  C_2\to \tfrac{3}{4}   \left(C_1\mp\sqrt{C_1 C_3} \right),\, C_4\to \tfrac{1}{4}  \left(-C_1\pm 4 \sqrt{C_1 C_3} -C_3\right),\, C_5\to \tfrac{3}{4}  \left(\mp\sqrt{C_1 C_3} +C_3\right)\,.
\end{eqnarray}}\\
Thus we see that in case (a) the boost $K$ generator restores and the symmetry is enhanced to a Galilean group.\footnote{In our convention, the Galilean group is generated by translations, rotations and boosts. One could also add in a dilatation generator, but the particle number will not be conserved under this scaling transformation for $z\neq 2$. Therefore in this paper we do not include this dilatation generator to be part of the Galilean group and consider it as part of the extension to the Schr\"{o}dinger group at $z=2$.} For another choice of the parameters (b), the scaling symmetry is respected. Furthermore, it is possible for some other choices of parameters (c), (d) that a linear combination of boost and scaling becomes a symmetry. But it is impossible that both of them become symmetry simultaneously; there are at most 4 generators in the symmetry of the solution. 

The solutions (a), (b) and (c) have different boundary behavior and hence are different physical solutions. While in case (d) the parameter $C_1$ is a gauge parameter that relates the solutions (d) to (c).  

We then consider the symmetries of the spin-4 metric like fields.  Astoundingly, the previously determined symmetries of the metric and spin-3 metric like fields are all symmetries of the spin-4 metric like field as well. This is very likely to be a consequence of the fact that in the frame like field $W$, only spin-3 components of the higher spin fields are turned on; even though the spin $s>3$ metric like fields are non-vanishing, they do not carry new physical information.\footnote{We thank Wei Li for a discussion on similar situations in 3D.} Therefore we believe the symmetries we have found previously are symmetries of the full solution that we have constructed.

\paragraph{Global internal symmetry}
Global symmetry of a vacuum solution to the Vasiliev equation can be extracted from the equation
\begin{equation}\label{gaugeequation}
d\epsilon(y|x) +[W,\epsilon(y|x)]_*=0\,,
\end{equation}
which determines how does a given symmetry parameter $\epsilon_0(y)$ at any fixed spacetime  point extend to a small neighborhood around this point. 
Since $W$ is a solution to the flatness equation, it is always possible to rewrite the vacuum solution in the form of a pure gauge in this neighborhood \cite{Vasiliev:1999ba,Giombi:2010vg,Bolotin:1999fa}.
\begin{equation}
W= g^{-1}(y|x)* dg(y|x)\,.
\end{equation}
The solution to the equation \eqref{gaugeequation} in this gauge can be trivially solved as \begin{equation}
\epsilon(y|x) =g^{-1}(y|x) * \epsilon_{0}(y)* g (y|x)\,,
\end{equation}
where $\epsilon_{0}(y)$  
does not depend on spacetime coordinates and fully determines the global (internal) symmetry. It is concluded in \cite{Lei:2015ika} that the symmetry of Schr\"{o}dinger higher spin solution in 3D Chern-Simons theory is just $SL(N,R) \times SL(N,R) $ by applying the gauge function method above. In the current higher dimensional case, one could also conclude that $\epsilon_0(y)$  exhausts the whole Vasiliev higher spin symmetry group. 
 
\paragraph{An interesting fact}
The Schr\"{o}dinger isometry algebra is actually a subalgebra of the global symmetry algebra discussed above. 
One way to realize this embedding is 
\bal \nn
& P=i T_{t x}\,,\qquad K=i T_{r x}+i T_{ \hat{d} x}\,,\qquad  D=-i T_{t x}+i T_{r \hat{d}}\,,\\ \nn
&  M=-i T_{rt}+iT_{\hat{d}t}\,,\qquad  C=\frac{i}{\sqrt{2}} (T_{rx}-T_{ \hat{d}x})\,, \qquad H=\frac{i}{\sqrt{2}} (T_{\hat{d}t}+T_{rt})\,,
\eal
where $\mu,\nu=\{t,x,\xi,r\}$, and 
\bal
T_{\mu \hat{d}} = \frac{1}{4} (\sigma_{\mu})_{\alpha \dbeta} y^{\alpha} \bar{y}^{\dbeta} =-T_{\hat{d}\mu }\,,\quad 
T_{\mu \nu} = \frac{1}{8}((\sigma_{\mu\nu} \epsilon)_{\alpha \beta} y^{\alpha}y^{\beta} +(\epsilon \bsigma_{\mu\nu})_{\dalpha \dbeta} \bar{y}^{\dalpha} \bar{y}^{\dbeta})\,.
\eal
The set $\{C,D,H,K,M,P\}$ generates the Schr\"{o}dinger symmetry group $Sch(1)$.

\subsection{Solutions with other scaling factors}

As we have mentioned in the introduction, $z=2$  Schr\"{o}dinger spacetime has a larger isometry group than  Schr\"{o}dinger spacetime with $z \neq 2$. To demonstrate that our construction is universal for all integer $z$ rather than merely a result of the larger symmetry group at $z=2$, we have also constructed the $z=3$ Schr\"{o}dinger spacetime in a similar way. The $z=3$ Schr\"{o}dinger spacetime turns out to be supported by spin-$4$ fields in the $t$ direction. We spare the reader from the tedious expression since it is not particularly illuminating. From the construction, we find explicitly that the back-reaction of spin-$s$ fields ``deforms" $AdS_4$ to  Schr\"{o}dinger spacetime in 4D with $z=s-1$.

A general spin-$s$ field $
W_{(2s-2)}=\left\{\omega_{\alpha_1 ... \alpha_{2s-2}}, \ ..., \ \omega_{\dalpha_1 ... \dalpha_{2s-2}} \right\}
$
has $N_s=\dfrac{s}{3}(4s^2-1)$ independent components, which is the same as the number of independent equations in \eqref{z=2Schrsolution4D}. 
 In other words, if one specifies a group of parameters as ``boundary conditions" of the differential equations, 
all the components of master field $W$ can be uniquely determined. Furthermore, if this group of parameters can be fixed from a given set of generalised vielbein, as in our spin-3 example, there is no degeneracy problem. This property can only be checked case by case.

\subsection{RG flows}

In the previous sections, we have considered solution to the Vasiliev equation that corresponds to spacetime with Schr\"{o}dinger isometry. These solutions are derived by turning on higher spin fields with one given spin. One immediate question is what if we turn on fields with different spins in a similar manner.\footnote{We thank Matthias Gaberdiel for pointing this direction to us.}

From the above construction, we notice that the higher spin fields only enter equation \eqref{4DSHCRmaster2} and hence fields with different spins are in general independent to each other. Therefore, the general solution with different higher spin fields turned on is simply a linear combination of the previous solutions where only one single higher spin field is turned on. Thus the general solution gives the following metric
\bal 
ds^2=(\sum_{i=i_{\text{min}}}^{i_\text{max}} \frac{f_i}{r^{2i-2}}) dt^2 + \frac{2dtd\xi +dr^2 +dx^2}{r^2}\,,
\eal
where the index $i_{\text{min}}$ and $i_{\text{max}}$ are the minimal and maximal spins we have turned on in the $t$-direction.
The number of independent parameter $f_i$ agrees with the number of different higher spin fields. Higher spin Fronsdal fields can be similarly determined.

Geometrically, these solutions interpolate between Schr\"{o}dinger-like geometries with different dynamic exponents.  This can be easily verified not only for the metric but also the higher spin Fronsdal fields. The existence of this RG type  solution is due to the presence of higher spin fields, as well studied in the pure AdS case \cite{Ammon:2011nk, Peng:2012ae}.

From the dual field theory point of view, these solutions correspond to RG flows between $U(N)$ models with different deformations, which we discuss in detail in Section \ref{fty}, resembling the RG flows between different Landau-Ginzburg models or minimal models in 2D. In the cases where the solutions respect the scaling symmetry, the dual RG flow is also interesting since that relates theories where the time direction scales differently.

\subsection{Linearised scalar equations}
We have discussed new exact ground-state solutions to Vasiliev equation \eqref{4Dflatequation} in 4D. In this section, we consider the motion of the scalar fields. We find the linearised scalar equation of motion on this Schr\"{o}dinger background to be ``deformation" of the free Klein-Gordon equation with extra radius-dependent source due to the spin-3 fields. Explicit calculations of correlation functions are left for future work \cite{LeiPeng}. 

The ground state configuration is 
\bal
W_0= W_2 +W_3\,, \qquad B_0=0\,, \qquad S_0=dz^{\alpha}\,z_{\alpha}+d\bar{z}^{\dot{\alpha}}\,\bar{z}_{\dot{\alpha}}\,.
\eal
Linearized perturbation around it means
\begin{equation}
W=W_0(x|Y) +{W}_1(x|Y,Z); \qquad S=S_0+S_1(x|Y,Z); \qquad B=B_1(x| Y ,Z)\,.
\end{equation}
The linearised Vasiliev equations are \begin{eqnarray} \nn
&& D_0 {W}_1 =0 \,, \\ \nn
&& \tilde{D}_0 B_1=0 \,, \\ \nn
&& d_Z {W}_1 +D_0 S_1 =0 \,, \\ \nn
&& d_Z S_1 =e^{i\theta_0} B_1*K dz^2 +e^{-i\theta_0} B_1*\bar{K} d\bar{z}^2\,, \\ \label{linearisationequation}
&& d_Z B_1=0 \,,
\end{eqnarray}

where $\theta_0$ is a parameter corresponding to the parity of the scalar. 
From the last equation of \eqref{linearisationequation}, $B_1$ is independent of $Z$. Then higher spin fields are
$$
C(x|Y)=B|_{Z=0}=B_1(x|Y)\;. 
$$
We consider the linearised equation of the scalar,
\begin{equation}
\tilde{D}_0 C= dC+W_0*C-C*\pi({W}_0)=0\,,
\end{equation}
where $\pi({W}_0(y,\bar{y}))=W_0(-y,\bar{y})$, which flips the sign of the coefficients of any odd number of ${y}$ oscillator. Therefore, it is useful to separate our background spin-3 field $W_3$ into two pieces that has even or odd number of (un)barred oscillators, namely $W_3=W^e_3+W^o_3$, respectively. In the direction other than $t$, the background gauge field is the same as that of the AdS spacetime, while in the $t$ direction, there are spin-3 fields turned on. Therefore we have 
\begin{equation}
\nabla^L_\mu C(x|Y)+2 (e_\mu)_{\alpha \dbeta}y^\alpha \bar{y}^{\dbeta}C(x|Y)+2 (e_\mu)^{\alpha \dbeta} \partial_{y^{\alpha}} \partial_{\bar{y}^{\dbeta}}C(x|Y)= N_3 \delta_{\mu t}\,,
\end{equation}
where \bal N_3 = N_3^e+N_3^o= -\big( [W^e_3,C(x|Y)]+\{W^o_3,C(x|Y)\}\big)\,.\label{n3}\eal

Following the notation of \cite{Giombi:2012ms}, $C^{(n,m)}$ represents the terms in $C(x,Y)$ of degree $n$ in $y$  and degree $m$ in $\bar{y}$. Scalar fields are contained in $C^{(n,n)}$. The $C^{(0,0)}$ equation receives higher spin corrections 
\begin{equation}\label{00piece}
\partial_\mu C(x)^{(0,0)}+2 (e_\mu)^{\alpha \dbeta} \frac{\partial}{\partial y^\alpha}\frac{\partial}{\partial \bar{y}^{\dbeta}}C^{(1,1)}
=N_3^o \delta_{\mu t}\,,
\end{equation}
where the commutator term in \eqref{n3} vanishes and hence we only keep the odd piece $N^o_3$ of $N_3$. Similarly, for the $C^{(1,1)}$ equation, we have
\begin{equation}\label{11piece}
\nabla^L_\mu C^{(1,1)}+2 (e_\mu)_{\alpha \dbeta}y^\alpha \bar{y}^{\dbeta}C^{(0,0)}+2 (e_\mu)^{\dbeta\alpha} \frac{\partial}{\partial y^\alpha}\frac{\partial}{\partial \bar{y}^{\dbeta}}C^{(2,2)}
=N_3^e \delta_{\mu t}\,.
\end{equation}
Expanding the above equation in component form, \bal C^{(1,1)}=C^{(1,1)}_{\alpha\dot{\beta}}y^\alpha \bar{y}^{\dot{\beta}}\,,\qquad  C^{(2,2)}=C^{(2,2)}_{\alpha\rho,\dot{\beta}\dot{\tau}}y^\alpha y^\beta\bar{y}^{\dot{\beta}}\bar{y}^{\dot{\tau}}\,,\eal we get 
\bal
\nn & C^{(1,1)}_{\gamma \dkappa}=4(e^\mu)_{\gamma \dkappa}\partial_\mu C(x)^{(0,0)}-   4 N_3^o  (e^t)_{\gamma \dkappa}\,,\\
& (e^\mu)^{\dkappa \gamma}\nabla^L_\mu C^{(1,1)}-\frac{1}{4}y^\gamma \bar{y}^{\dkappa}C^{(0,0)}-\frac{1}{2}\epsilon^{\dkappa \dbeta} \epsilon^{\gamma \alpha}  y^\rho \bar{y}^{\dtau} C^{(2,2)}_{\alpha \rho, \dbeta \dtau}
=(N_3^e)_{\rho \dtau} y^\rho \bar{y}^{\dtau} (e^t)^{\dkappa \gamma}\,.\label{0011}
\eal
Eliminating $C^{(2,2)}$ term by acting the equation with $\partial_{y^{\alpha}} \partial_{\bar{y}^{\dbeta}}$, we get our final deformed Klein-Gordon equation 
\bal
-\frac{1}{2}(\nabla_\mu\partial^\mu+2) C^{(0,0)}= (e^\mu)^{\dkappa \gamma} (e^t)_{\gamma\dkappa} (  4 \partial_\mu N_3^o  )+(N_3^e)_{\gamma\dkappa}  (e^t)^{\dkappa \gamma}\,.
\eal
This equation is simply the normal Klein-Gordon equation sourced by the known function $N_3$. We can solve it as the motion of the scalar under a classical potential $N_3$ due to the spin-3 fields. We will report the detailed analysis in the near future \cite{LeiPeng}.

\section{Schr\"{o}dinger solution in general dimension}
Our construction is in fact applicable in any dimension where the Vasiliev higher spin theory is defined. One interesting example is  Schr\"{o}dinger solutions in the $6D$ higher spin gravity since the isometry group of the metric in that case is $Sch(3)$, which is the same symmetry governing the $3D$ unitary fermion theory that can be used to describe the cold atom system \cite{Son:2008ye, Balasubramanian:2008dm, Bekaert:2011cu, Bekaert:2011qd}. 

We use the vectorial formalism \cite{Vasiliev:2004cp, Bekaert:2005vh} for the construction in general $d$-dimension. In particular, we have the generators of the AdS spacetime isometry
\begin{equation}
T^{AB}=-T^{BA}=\epsilon^{ij}Y^A_iY^B_j\,,
\end{equation}
where $i,j=1,2$ are the $sp(2)$ indices and $A,B=t,\x,x_1,\ldots,x_{d-2},r,\hat{d}$ are the (extended) spacetime index. The $\hat{d}$ is an auxiliary direction with negative signature. The master gauge field $W$ encodes the vielbein and spin connections
\begin{align}
W\supset W^{(2)}=W_{ab} T^{ab}\,, \qquad  e^a=W^{a\hat{d}}\,,\quad w^{ab}=W^{ab}\,.
\end{align}
The AdS solution is simply
\begin{align}
e_a=W_{a\hat{d}}=\frac{\delta_{a\mu} dx^\mu}{r}\,,\qquad  w_{ab}=-w_{ba}=\frac{ \delta_{ar}\delta_{b\mu}-\delta_{b r}\delta_{a\mu}}{r}\,.
\end{align}

One special feature of the vectorial formalism is the requirement of factoring out the ideal that is generated by the $sp(2)$ generators \cite{Vasiliev:2004cp, Bekaert:2005vh}. This process puts the set of equations on mass-shell.  However, this process of factoring out the ideal can be farely complicated in general dimension. We have constructed solutions in 4D and 6D vector like formalism by turning on spin-3 fields in the $t$-direction, but the solutions are still off mass-shell at the current stage. We will put complete analysis in future work \cite{LeiPeng}.

As a further evidence, we can realize the Schr\"{o}dinger symmetry generators in terms of the vectorial oscillators
\bal 
\nn & M^{ij}=i T^{x_i\,x_j}\,,\qquad  P^i=i T^{t\, x_i}\,,\qquad  K^i=i T^{r\, x_i}-i T^{ \hat{d}\, x_i}=i T^{-\, x_i}\,,\\
\nn & C=\tfrac{i}{\sqrt{2}} (T^{r\,\x}-T^{ \hat{d}\,\x})=\tfrac{i}{\sqrt{2}} T^{-\,\x}\,,\quad H=\tfrac{i}{\sqrt{2}} (T^{\hat{d}\,t}+T^{r\,t})= \tfrac{i}{\sqrt{2}}  T^{+\,t}\,,\\
& D=-i T^{t\, \x}+i T^{r\, \hat{d}}\,,\qquad M=-i T^{r\,t}+i T^{\hat{d}\,t}=-i T^{-\,t}\,
\eal
where we have defined a second pair of light-cone coordinates $\pm=r\pm \hat{d}$. 
 
As in the 4 dimensional case, we expect the symmetry of the full solution to be the Galilean group, with possible mixing with scaling transformation.

\section{Dual field theory}\label{fty}
There are two parity invariant  Vasiliev theories on $AdS_4$ that are dual to bosonic \cite{Klebanov:2002ja} and fermionic \cite{Sezgin:2002rt,Sezgin:2003pt} versions of $O(N)$ or $U(N)$ vector models respectively (see e.g. \cite{Giombi:2012ms} for a review). Since our background can be explicitly check to be parity invariant and we can put the same two types of Vasiliev theories on our background, we believe that there should also be a bosonic and a fermionic version of the holographic dual field theories. 
In this section, we discuss such bosonic and fermionic field theories, which are simply free $U(N)$ field theories with spin-$3$ operators in the action that respect the same symmetries as the bulk solutions. They are valid candidates as the dual theories since our solutions in the bulk are constructed in the same manner by turning on spin-3 fields on top of the AdS geometry. The bulk scalar can have different boundary behaviors; the corresponding dual theories can be derived from the theories we proposed below by adding double-trace operators and flow to the IR/UV fix points. This is in parallel to the AdS case so we will not go into the details.

\subsection{Theories with Galilean symmetry}
We have shown that a class of our solutions respects the Galilean symmetry. The corresponding field theories on the boundary can   be obtained from a CFT by turning on extra spin-3 current operators
\begin{equation}
S= S_{\text{CFT}} + \int dt dx d\xi\Phi^{\mu\nu\rho}\mathcal{J}_{\mu\nu\rho}\,,
\end{equation}
that break the relativistic conformal symmetry but preserve the same symmetry as in the bulk.
In particular, the field theory should preserve the Galilean boost symmetry 
\bal
\xi' = \xi+vx-\frac{1}{2}v^2 t\,, \qquad
x' = x-vt\,, \quad \forall v\,.
\eal
As it is discussed in section \ref{sec:sym}, only $\Phi_{ttt}\sim\Phi^{\xi\xi\xi}r^{-6} \sim r^{-4}$ component is present in the bulk. Thus we propose that the dual bosonic field theory with the Galilean symmetry has the action 
\begin{equation}
S_B=\int dtd\xi d\vec{x}^{D-3} \Big(\pa_t\bar{\varphi}^a \pa_\x \varphi^a + \pa_\x\bar{\varphi}^a \pa_t \varphi^a +
\pa_{\vec{x}}\bar{\varphi}^a \pa_{\vec{x}} \varphi^a - \Sigma \bar{\varphi}^a \partial_{\xi}^3 \varphi^a) \,,\label{bosft}
\end{equation}
where $a$ is the $U(N)$ index and $\Sigma$ has mass dimension $-1$.
The equation of motion is \begin{equation}\label{Bosoneom}
H_B\,\varphi^a=\Big(2 \pa_t\pa_\x+\pa_{\vec{x}}^2 + \Sigma \partial_{\xi}^3  \Big) \varphi^a=0\,.
\end{equation}
It's easy to check that the equation is preserved by time translation $H$, momentum $\vec{P}$, non-relativistic mass $M$ and Galilean boost $\vec{K}$,  
i.e. the Galilean group.

Like the case of AdS holography, we expect another fermionic theory to be dual to the bulk higher spin gravity with a parity odd scalar. Following  the same reasoning as the bosonic theory \eqref{bosft}, it would be natural to propose a spin-3 ``deformed" fermionic free $U(N)$ theory, which can be defined by the following action
\begin{equation}\label{fermgal}
S_F =\int dtd\xi d\vec{x}^{D-3} \ (i\bpsi^a \Gamma^\mu \pa_\mu \psi^a  
-i{\Sigma} \bpsi^a   \Gamma^t\partial_\xi^2\psi^a)\,,
\end{equation}
where $\mu$ runs over all the spacetime indices and we have used the following definition of the $D$-dimensional gamma matrices in the lightcone coordinates 
\bal
\Gamma^t= \frac{1}{\sqrt{2}}(\Gamma^0+\Gamma^1)\,, \qquad \Gamma^\xi= \frac{1}{\sqrt{2}}(-\Gamma^0+\Gamma^1)\,, 
\eal
where $\Gamma^t, \Gamma^{\xi}$ satisfy \begin{equation}
\Gamma^t \Gamma^t = \Gamma^\xi  \Gamma^\xi =0; \qquad \{\Gamma^t ,\Gamma^\xi\} =2 I \,.
\end{equation}
In $3D$, a representation of these matrices can be chosen to be $ \Gamma^0 =\sigma_x \sigma_z\,, \Gamma^1 =\sigma_z\,, \, \Gamma^2 =\sigma_x $.  The equation of motion is \begin{equation}
H_F \,\psi^a= (\Gamma^\mu \partial_\mu -\Sigma \Gamma^t \partial_\xi^2) \psi^a=0\,.
\end{equation}
The action can be explicitly shown to be invariant under the Galilean symmetry group, which is not hard to understand since $H_B=H_F^2$. 

\subsection{Theories with non-relativistic scaling symmetry}
Another family of solutions with enhanced non-relativistic scaling symmetry has the following spin-3 fields \begin{equation}
\Phi_{ttt} = \frac{-3072  C_1^3}{r^6}\;, \qquad \Phi_{tt\xi} = -\frac{8C_1}{r^4}\;, \qquad \Phi_{txx} =\frac{8C_1}{r^4}\,.
\end{equation}
Note that $\Phi_{ttt} $ has distinct dimension from the other terms, which makes writing down an action for the dual field theory more challenging since we expect a spin-3 conserved current to have dimension $\Delta =4$ in a 3D CFT. In the language of field theory, this difficulty is that the $\bar{\psi}^a\partial_{\xi}^3\psi^a$ component in the higher spin currents 
does not respect the non-relativistic scaling.
Therefore,  
there seems to be no straightforward way to embed non-relativistic scaling symmetry into higher spin symmetry in the current construction unless sever modification is made. This result is somewhat consistent with the result obtained from 3D Chern-Simons theory. However, the $r^{-6}$ power hints on the possibility of terms that are not components of the higher spin currents, such as the multi-trace operator $(\bar{\psi}^a\partial_{\xi}\psi^a)^3$, to appear in the action.
This term is possible in light of the non-linear nature of bulk higher spin field $\phi_{ttt}$. Notice that this multi-trace operators can be understood, at least in the large $N$ limit, to ``run" the theory to some UV conformal fixed point, as in the more familiar AdS case \cite{Klebanov:2002ja, Giombi:2011kc, Chang:2012kt, Aharony:2012nh, Zayas:2013qda}.

On the other hand, there are known examples where the non-relativistic scaling is incorporated into the symmetry of the theory; these are the well known theories with the Schr\"{o}dinger symmetry, which is an extension of the Galilean symmetry by a non-relativistic dilatation and a special conformal transformations. One can construct such theories 
by ``deforming" a free CFT with spin-3 current operators
\begin{itemize}
	\item Bosonic
	\begin{equation}
	S_{\text{schr}}^B= \int dx^{D-1} \Big(\pa_t\bar{\varphi}^a \pa_\x \varphi^a + \pa_\x\bar{\varphi}^a \pa_t \varphi^a +\pa_{\vec{x}}\bar{\varphi}^a \pa_{\vec{x}} \varphi^a -\Sigma \bar{\varphi}^a (\partial_{\xi}\partial_x^2+ 2\partial_t \partial_\xi^2) \varphi^a) \,,\label{bossch}
	\end{equation} 
	\item Fermionic
	\begin{equation}
	S_{\text{schr}}^F= \int dx^{D-1} \ (i\bpsi^a \Gamma^\mu \pa_\mu \psi^a  
	-i\Sigma \bpsi^a \Gamma^t  (\partial_{\vec{x}}^2+2\partial_t \partial_\xi)\psi^a)\,.\label{fersch}
	\end{equation}
\end{itemize}
It is easy to check directly that the above theories have the Schr\"{o}dinger symmetry. There is a qualitative way to understand the presence of this symmetry. Taking the action \eqref{bossch} as example, the corresponding equation of motion reads
\bal 
(1+\Sigma\pa_\xi)(2\pa_t\pa_\xi+\pa_{\vec{x}}^2)\varphi^a=0\,.
\eal 
From which we see explicitly that the symmetry of this equation of motion are the subset of the symmetries of the Klein-Gordon equation 
that further commute with the $1+\Sigma \pa_\xi$ factor. This subset is the centralizer of $\pa_\xi+\tfrac{1}{\Sigma}$, which is nothing but the Schr\"{o}dinger group by construction of the light-cone reduction  \cite{Maldacena:2008wh,Bekaert:2011cu}. Moreover, since the $\pa_\xi$ plays the role of the non-relativistic mass generator $M$, the meaning of $\Sigma$ in the action \eqref{bossch} is then the (minus) inverse mass in the theory. This interpretation also holds in action with only Galilean symmetry, namely \eqref{bosft}. A similar argument applies to the fermionic action. 
This action then suggests that $\phi_{ttt} $ term vanishes in the bulk and boundary values of $\phi_{txx} $ and $\phi_{tt\xi}$ differ by a factor of 2. We do not observe this in the Schr\"{o}dinger solutions we have constructed. It would be interesting to construct a solution of the Vasiliev higher spin theory that is dual to the above Schr\"{o}dinger invariant field theory. Furthermore, it is interesting to see if there are higher spin solutions with the Galilean conformal symmetry \cite{Bagchi:2009my}.

\section{Discussion}

In this paper, we have constructed  solutions of the Vasiliev higher spin theory with Galilean symmetry in general dimensions. We show that the spacetime symmetry group can be the Galilean group or a non-relativistic scaling symmetry group. We further conjecture a bosonic and a fermionic field theory that could be dual to the type-A and type-B Vasiliev theories living on the  Schr\"{o}dinger background that we have constructed. The difference of the two types is only visible when considering the motion of the scalar probes, whose linearised equation of motion is also derived. Therefore the immediate next step is to consider correlation functions of the bulk higher spin system on the  Schr\"{o}dinger background and in the dual field theories we have proposed. This would provide another piece of strong evidence of whether our proposal is sensible or not.
This is currently under investigation \cite{LeiPeng}.

One general property of the higher spin gravity is that some usual geometric quantities such as event horizon might not remain physical observable in higher spin theory \cite{Gutperle:2011kf, Ammon:2011nk}. In fact, there are even different ways of identifying the gravity sector in a given higher spin system, which leads to interesting observations \cite{Ammon:2011nk, Peng:2012ae}. 
One proposal due to these special properties is the resolution of black hole or cosmic singularities by performing higher spin gauge transformations \cite{Castro:2011fm, Burrington:2013dda}. It is argued in \cite{Lei:2015ika} that this method cannot be used to resolve IR tidal force singularity in $3D$ Lifshitz \cite{Horowitz:2011gh, Copsey:2010ya} and $1<z<2$ Schr\"{o}dinger spacetime \cite{Blau:2009gd}  because degeneracy problem makes the spacetime interpretation problematic. However it is possible to construct $z=2$ $3D$ Lifshitz spacetime by dimensional reduction. One can show that if one adds a constant one-form $\eta= \eta_t dt$ to the AdS gravitational connection 
\begin{equation}
e= e_{\alpha \dbeta} y^{\alpha} \bar{y}^{\dbeta} + \eta; \qquad \omega= \omega_{\alpha \beta} y^{\alpha} y^{\beta} +\omega_{\dalpha \dbeta} \bar{y}^{\dalpha} \bar{y}^{\dbeta}\,,
\end{equation}
the master field $W$ still solves Vasiliev equation. It turns out that the corresponding metric represents the $z=0$  Schr\"{o}dinger spacetime \begin{equation}
ds^2= -\eta_t^2 dt^2 +\frac{2dt d\xi +dr^2 +dx^2}{r^2}\,.
\end{equation}
To proceed, we use the fact that $D-1$ dimensional $z=2$  Lifshitz spacetime emerges from $z=0$ Schr\"{o}dinger spacetime in $D\ge 4$ dimension by dimensional reduction in the $t$ direction \cite{Chemissany:2011mb, Chemissany:2012du}. Those $3D$ Lifshitz spacetimes are solutions of Einstein equation with supporting matter fields and therefore safe from degeneracy problem in higher spin theory \cite{Lei:2015ika}. One may be able to study how higher spin transformation operates on the Lifshitz geometry, and understand the physical meaning of IR singularity  \cite{Andrade:2014bsa}. The recent development of invariant functional \cite{Vasiliev:2015mka} for the Vasiliev theory could help make progress in this direction.

Last but not the least, it would be interesting to know whether Schr\"{o}dinger black hole solution exists in $4D$ Vasiliev theory. The known higher spin solution in 3 dimension \cite{Didenko:2006zd, Gutperle:2011kf}, the charged black hole solution with asymptotic  Schr\"{o}dinger geometry \cite{Herzog:2008wg, Maldacena:2008wh,Adams:2008wt} together with the reformulation of $AdS_4$ Kerr black hole solution into the unfolding formalism \cite{Didenko:2008va} hint on possibility of finding black hole solutions with asymptotic Schr\"{o}dinger geometry in higher spin theory. We will leave this for future work.

\section*{Acknowledgement}
We are grateful to Matthias Gaberdiel, Simone Giombi, Daniel Grumiller, Wei Li, Simon Ross and Stefen Theisen for useful discussions. We thank Eurostring 2015 in Cambridge University where the authors meet with each other and this project started. The research of C.P.\ is supported by a grant of the Swiss National Science Foundation, also partly by the NCCR SwissMAP.

\appendix

\section{Higher spin algebra in $D=3$}
\label{lonestar}
We will follow the notation in \cite{Gaberdiel:2012uj, Gaberdiel:2013jca}. The higher spin algebra $hs[\lambda]$ generator $V^s_m$ are defined to be 
\begin{equation}\label{defineV}
V_m^s = (-1)^{s-1-m} \frac{(s+m-1)!}{(2s-2)!}\left[ \underbrace{V_{-1}^2,... [V_{-1}^2,[V_{-1}^2}_{s-m-1 \text{ terms}},(V_1^2)^{s-1}]]\right]\,,
\end{equation}
where $$
V^2_1 =L_1\,, \quad V^2_0 =L_0\,, \quad V^2_{-1}= L_{-1}\,.
$$
If $\lambda= N$, the algebra is truncated to $sl(N)$ and all the $s>N$ generators can be removed. The lone star product between generators has a closed form
\begin{equation}
V_m^s * V_n ^t = \frac{1}{2} \sum_{u=1}^{s+t-|s-t|-1} g_{u}^{st}(m,n;\lambda) V_{m+n}^{s+t-u}\,,
\end{equation}
with 
$$
 g_{u}^{st}(m,n;\lambda) =(\frac{1}{4})^{u-2} \frac{1}{2(u-1)!}\phi_u^{st}(\lambda) N_u^{st}(m,n)\,,
$$
where $$
N_u^{st}(m,n) =\sum_{k=0}^{u-1}(-1)^k\binom{u-1}{k}[s-1+m]_{u-1-k} [s-1-m]_k [t-1+n]_{k} [t-1-n]_{u-1-k}\,,
$$
$$
\phi_u^{st}(\lambda)= {_4 F_3} \left[ 
\begin{matrix}
\frac{1}{2}+\lambda, \frac{1}{2}-\lambda, \frac{2-u}{2},\frac{1-u}{2}\\
\frac{3}{2}-s, \frac{3}{2}-t,\frac{1}{2}+s+t-u
\end{matrix}
\mid1
\right]\,.
$$
Here $[a]_n=a(a-1)...(a-n+1)$ are the descending Pochhammer symbol. The commutator of two generators are defined as \begin{equation}
[X,Y]= X*Y -Y*X\,.
\end{equation}
$V_m^s$ transforms in the $(2s-1)$ dimensional representation of $sl(2)$ Lie algebra
\begin{equation}\label{usefulcommutator}
[V_m^2, V_m^s]=(-n+m(s-1))V_{m+n}^s\,,
\end{equation}
which is also one of the useful formulas used in verifying Schr\"{o}dinger solution. 
The trace of lone star product is defined to be \begin{equation}\label{trace}
\text{tr}(X*Y) = X* Y\mid_{V^s_m=0,s>0}\,.
\end{equation}
The relation with the oscillator realization is via the identification 
\begin{equation}
V^2_1 = \frac{1}{2} T_{11}\,, \qquad V^2_0 = \frac{1}{2}T_{12}\,, \qquad V^2_{-1}= \frac{1}{2} T_{22}\,.\label{sl2id}
\end{equation}
Other higher spin generators $V_m^s$ are related to $T_{\alpha \beta}$ via equation \eqref{defineV}.

\bibliographystyle{JHEP}
\bibliography{higherspin}

\end{document}